\documentclass[pra,amsmath,amssymb,twocolumn]{revtex4-1}
\usepackage[usenames]{color}
\usepackage{amssymb,hyperref}
\usepackage{grffile}
\usepackage[pdftex]{graphicx}
\usepackage{amsmath, amstext, amssymb, amsfonts, amsxtra}
\usepackage[tbtags]{mathtools}
\usepackage{textcomp}
\usepackage{xspace}
\usepackage{icomma}
\usepackage[short]{datetime}

\def \etal{{\it {et al.}}}
\def \ell{{d}}

\def \fockket{{\ket{\nnot}}}

\def \nex{n_{\mbox{\small ex}}}

\def \k{{k}}

\def \nnot{{\bar{n}}}

\def \ketinit{\ket{\psi_{\text{\tiny init}}}}

\def \av#1{{\langle#1\rangle}}

\renewcommand{\hat}{{}}

\newcommand{\bra}[1]{\ensuremath{\langle #1 \vert}\xspace}%
\newcommand{\ket}[1]{\ensuremath{\vert #1 \rangle}\xspace}%
\newcommand{\avg}[1]{\ensuremath{\langle #1 \rangle}\xspace}%

\newcommand{\plus}{{+}}
\newcommand{\minus}{{-}}
\newcommand{\plusminus}{{\pm}}
\newcommand{\lproj}{{\mathcal{P}_j}}
\newcommand{\gproj}{{\boldsymbol{\mathcal{P}}}}
\newcommand{\com}[2]{[{#1},{#2}]}
\newcommand{\anticom}[2]{\{{#1},{#2}\}} 
\newcommand{\Jb}{\mathcal{J}}

\newcommand{\vinfty}{\tilde{J}}

\def \double{{\ket{\bar{n}+1}}}%
\def \hole{{\ket{\bar{n}-1}}}%
\def \single{{\ket{\bar{n}}}}%

\def \figpreamble{(Color online) }

\newcommand{\unige}{D\'epartement de Physique Th\'eorique, Universit\'e de Gen\`eve, 1211
  Gen\`eve, Switzerland}%
\newcommand{\mpq}{Max-Planck-Institut f\"ur Quantenoptik, 85748 Garching, Germany}%
\newcommand{\cpht}{Centre de physique th\'{e}orique, \'{E}cole Polytechnique, CNRS, 91128 Palaiseau, France}

\begin{document}

\title{Propagation front of correlations in an interacting Bose gas }

\author{Peter Barmettler$^1$}%
\author{Dario Poletti$^1$}%
\author{Marc Cheneau$^2$}%
\author{Corinna Kollath$^{1,3}$}%
\affiliation{$^1$\unige}%
\affiliation{$^2$\mpq}%
\affiliation{$^3$\cpht}

\date{\today}

\begin{abstract}
  We analyze the quench dynamics of a one-dimensional bosonic Mott insulator and focus on the time
  evolution of density correlations. For these we identify a pronounced propagation front, the
  velocity of which, once correctly extrapolated at large distances, can serve as a quantitative
  characteristic of the many-body Hamiltonian. In particular, the velocity allows the
  weakly interacting regime, which is qualitatively well described by free bosons, to be distinguished
  from the strongly
  interacting one, in which pairs of distinct quasiparticles dominate the dynamics. In order to
  describe the latter case analytically, we introduce a general approximation to solve the
  Bose--Hubbard Hamiltonian based on the Jordan--Wigner fermionization of auxiliary particles. This
  approach can also be used to determine the ground-state properties. As a complement to the
  fermionization approach, we derive explicitly the time-dependent many-body state in the
  non-interacting limit and compare our results to numerical simulations in the whole range of
  interactions of the Bose--Hubbard model.
\end{abstract}

\maketitle

\section{Introduction}

In the past two decades, progress in atomic physics, quantum optics, and the nanosciences has
propelled quantum many-body theory to meet new challenges. 
It is indeed now possible to engineer systems that are concrete realizations of some paradigmatic models, which
were once introduced to grasp fundamental properties of more complex materials.  New frontiers thus
have to be explored, among which the dynamics of these isolated quantum models far from equilibrium
is probably the least well understood and one of the most exciting.

One of the fundamental questions that has to be addressed is how correlations propagate in these
systems.  The Schr\"odinger equation allows in principle for correlations between distant points to
build up in arbitrary short times \cite{Bravyi2006}.  This contrasts with relativistic quantum field
theories, where physical effects cannot propagate faster than the speed of light and causality
relations between two points in space-time can exist only if one lies within the light-cone of the
other. In a seminal work \cite{Lieb1972}, Lieb and Robinson however showed that non-relativistic
quantum many-body systems can still exhibit some sort of locality: In generic one-dimensional spin
models with finite-range interactions, the propagation of correlations appears to be bounded by an
effective light cone, outside which correlations are exponentially suppressed. Here the role of
the speed of light is played by a velocity which is an intrinsic property of the many-body
Hamiltonian. The existence of so-called Lieb--Robinson bounds has many far-reaching
implications. For example, they make it possible to simulate on classical computers the ground state
properties as well as the dynamical evolution of such quantum systems \cite{Eisert2010,
  Hastings2009, Schollwock2011, Enss2012}. They also provide a general link between the presence of
a finite spectral gap and the existence of a finite correlation length in the ground state of
certain lattice systems \cite{Nachtergaele2006, Hastings2006, Nachtergaele2010,
  Nachtergaele2011}. However, the extent to which Lieb--Robinson bounds can be generalized beyond
spin systems remains an open question. Proofs or evidence for the existence of such bounds have
indeed been reported in various systems, ranging from harmonic chains to the Bose--Hubbard model
\cite{Calabrese2006, Calabrese2007, Cramer2008, Nachtergaele2009, Lauchli2008, Manmana2009}.  But it
is also possible to construct models in which the propagation velocity of correlations is explicitly
unbounded \cite{Eisert2009}.

Dynamical properties of correlations in a closed system can be probed by studying the time evolution
following a sudden change of parameter in the Hamiltonian, a situation referred to as a \emph{quantum
  quench}.
The quench has a particular appeal in the context of ultracold gases in optical lattices as the
relevant parameters in the Hamiltonian governing these systems can be easily varied in time
\cite{Bloch2008}.  In addition to the existence of an effective light cone, it was discovered in recent
theoretical studies that the time evolution of correlations in quenched systems is characterized by
a pronounced propagation front \cite{Lauchli2008, Cheneau2012, Calabrese2006, Calabrese2007,
  Cazalilla2006, Iucci2009, Sotiriadis2010, Igloi2000, Cubitt2008, Manmana2009, Mathey2010,
  Goth2012}. As for a Lieb--Robinson bound, the velocity at which this front propagates
can involve a broad range of the
spectrum of the Hamiltonian since the system is far from equilibrium. This makes
the understanding of this feature particularly challenging: Covariant low-energy effective theories
would provide a natural description \cite{Calabrese2006, Calabrese2007, Cazalilla2006, Iucci2009,
  Sotiriadis2010}, but, due to the presence of high-energy excitations, realistic lattice models
\cite{Lauchli2008, Cheneau2012, Calabrese2006, Calabrese2007, Igloi2000, Amico2004a, Rigol2006,
  Chiara2006, Cubitt2008, Barmettler2008, Lauchli2008, Manmana2009, Mathey2010, Barmettler2010,
  Rieger2011, Goth2012, Natu2012} show a much richer behavior than their corresponding field
theories. Gaining more insight into the non-equilibrium properties of quantum systems thus urges the
development of new effective models.

In a recent work \cite{Cheneau2012}, the propagation of correlations in a quantum many-body system
was studied both theoretically and experimentally in a one-dimensional bosonic gas in an optical
lattice and a propagation front could be clearly identified. The observed behavior was interpreted
using an exactly solvable effective model derived from the Bose--Hubbard Hamiltonian and describing
non-interacting fermionic quasiparticles. The key idea behind this model is to use a Jordan--Wigner
transformation to cure some of the problems inherent to the slave-boson methods proposed previously
\cite{Altman2002, Huber2007}. In the present article, we describe this approach in more detail
and use it to derive the ground state as well as the quench dynamics in the Mott-insulating phase. We
show that its predictions are quantitatively correct in a regime of strong and intermediate
interactions. Our model, being exactly solvable, allows us to explore the time evolution of the
system at long times, and we can show that the velocity of the propagation front exhibits a generic
scaling behavior.  Using numerical simulations, we find that this behavior holds in all interaction
regimes, down to the non-interacting limit of free bosons, where explicit solutions are
available. The asymptotic value of the velocity of the propagation front, which strongly differs
between the strongly and the weakly interacting limits, can be used to characterize the crossover
between these two regimes.

This article is organized as follows: In Sec.~\ref{sec:model} we present the model that we will
study; in Sec.~\ref{sec:fermion_approach} we carry out the fermionization procedure and derive
general relations that enable the calculation of equilibrium (Sec.~\ref{sec:equ_prop}) and
non-equilibrium (Sec.~\ref{sec:timeevolution}) properties. The velocity of the propagation front at
weak and strong interactions is analyzed in Sec.~\ref{sec:prop-quas}.  In Sec.~\ref{sec:conclusions}
we present our conclusions.

\section{One-dimensional system of bosonic atoms in an optical lattice}
\label{sec:model}
In this work we consider a one-dimensional system of bosonic atoms in an optical lattice. If the
lattice is deep enough, this system can be described by the one-dimensional single-band
Bose--Hubbard Hamiltonian:
\begin{align}
  \label{eq:bhHamiltonian}
  \hat H = \sum_{j} \left\{ -J \,(\hat a^{\dagger}_{j} \, \hat a_{j+1} + \text{h.\,c.}) +
    \frac{U}{2} (\hat n_{j}- \bar{n})^2 \right\} \; ,
\end{align}
where $\hat a_{j}$ and $\hat a^{\dagger}_{j}$ represent the annihilation and creation operators of a
bosonic atom at site $j$ and $\hat n_{j} = \hat a^{\dagger}_{j} \hat a_{j}$ counts the number of
atoms at that site. We use a lattice constant $a_{\text{lat}}=1$ and the system is considered to be
infinitely large and homogeneous. The kinetic part of the Hamiltonian is characterized by the
hopping amplitude $J$; the on-site interaction strength $U$ is related to the $s$-wave scattering
length. We work at fixed commensurate filling $\bar{n}$, where the model exhibits a quantum phase
transition between a superfluid phase at low interaction strengths $U/J<(U/J)_{\text c}$ and a
Mott-insulating phase at large interaction strengths $U/J>(U/J)_{\text c}$. At the specific filling
$\bar n=1$, the critical value is given by $(U/J)_{\text c} \sim 3.4$ \cite{Kashurnikov1996,
  Kuehner1998}.
The Bose--Hubbard model is non-integrable \cite{Kolovsky2004a, Kollath2010a} and exhibits complex
many-body properties; in particular, its non-equilibrium properties are far from being fully understood.

In order to benchmark the analytical approaches, we will
perform exact numerical simulations of model \eqref{eq:bhHamiltonian} by means of the density
matrix renormalization group (DMRG) \cite{White1992, Schollwock2005}, an algorithm based on matrix
product states \cite{Schollwock2011}. While the DMRG algorithm gives highly accurate results
for the ground state, time evolution \cite{Vidal2003, Daley2004, White2004} can be calculated only for
relatively short periods of time.

\section{Fermionization approach to the study of the Bose--Hubbard model}
\label{sec:fermion_approach}
In the following, we will describe in detail how the Bose--Hubbard model can be mapped onto an
effective model of non-interacting auxiliary fermions. The procedure consists of four main steps:
(i) The local Hilbert space is reduced to only three states and (ii) auxiliary bosonic operators are
introduced that allow switching between these states (Sec.~\ref{sec:auxiliarybosons}); (iii) the
auxiliary boson operators are fermionized by a Jordan-Wigner transformation
(Sec.~\ref{sec:fermionization}); (iv) a constraint on the fermionic operators is relaxed so that the
effective Hamiltonian becomes quadratic and can be diagonalized (Sec.~\ref{sec:unconstrained}).

\subsection{Auxiliary boson representation}
\label{sec:auxiliarybosons}
In the Mott-insulating phase and away from the critical point, the local density fluctuations around
the average filling $\nnot$ are limited. It is thus possible to truncate the local basis on a single
site $j$ to three states only: $\ket{\nnot+m}_j$, with $m=-1,0,1$.  Within this reduced basis, one can
represent the bare atomic operators $a_j^{(\dagger)}$ in terms of constrained auxiliary boson
operators $b^{(\dagger)}_{j,\sigma}$ with two flavors $\sigma=\pm1\equiv\plusminus$:
\begin{align}
  \hat a^\dagger_j=\sqrt{\nnot+1}\; \hat{b}_{j,\plus}^\dagger +\sqrt{\nnot} \;
  b_{j,\minus}\label{eq:sbmapping} \,.
\end{align}
The `$\plus$'-bosons correspond to excess particles: $\hat b_{j,\plus}^\dagger\single_j=\double_j$
, and `$\minus$'-bosons to holes: $\hat b_{j,\minus}^\dagger\single_j=\hole_j$. The local Fock state
$\ket{\nnot}_j$ represents the vacuum state of the auxiliary particles
$b_{j,\sigma}\ket{\nnot}_j=0$. Bosonic commutation relations are obeyed:
\begin{align}
  \com{b_{j,\sigma}}{b^\dagger_{j',\sigma'}} & = \delta_{j,j'}\delta_{\sigma,\sigma'}\,,\notag\\%
  \com{b^\dagger_{j,\sigma}}{b^\dagger_{j',\sigma'}}&= \com{ b_{j,\sigma}}{b_{j',\sigma'} }= 0\,,%
\end{align}
which allow for the unphysical situation of single sites being occupied by more than one auxiliary
boson. Therefore, the auxiliary operators have to fulfill the hardcore constraint
\begin{align}
  \left(\hat b^\dagger_{j,\sigma}\right)^2=\left(\hat b_{j,\sigma}\right)^2=0
  \label{eq:constraint1}
\end{align}
and double occupancies of different species need to be eliminated by imposing
\begin{align}
  \hat b^\dagger_{j,\plus}\hat b_{j,\plus}\hat b^\dagger_{j,\minus}b_{j,\minus}&=0\,.
  \label{eq:constraint2}
\end{align}
This representation in terms of doubly-flavored constrained bosons is slightly different from the
one used in slave-particle techniques \cite{AuerbachBook, Altman2002, Dickenscheid2003,
  Lu2006,Huber2007, Tokuno2011b}, in which one introduces one auxiliary operator for each of the
three local states $\ket{\nnot+m}_j$ and the number of auxiliary bosons per site is constrained to
be exactly one.

\subsection{Fermionization}
\label{sec:fermionization}
It is difficult to ensure the operator constraints \eqref{eq:constraint1} and \eqref{eq:constraint2} 
in general, and one often resorts to mean-field \cite{Dickenscheid2003, Lu2006, Tokuno2011b}
and perturbative \cite{Altman2002,Huber2007,Tokuno2011b} approximations.  In the special one-dimensional case,
however, it is possible to use Jordan--Wigner transformations \cite{Jordan1928, Batista2001} which
allow on the one hand for the exact treatment of the hard-core constraint (\ref{eq:constraint1}) and on
the other for suppression of local pairing of auxiliary particles.

\begin{figure*}[ht]
  \begin{center}
    \includegraphics[width=0.99\textwidth]{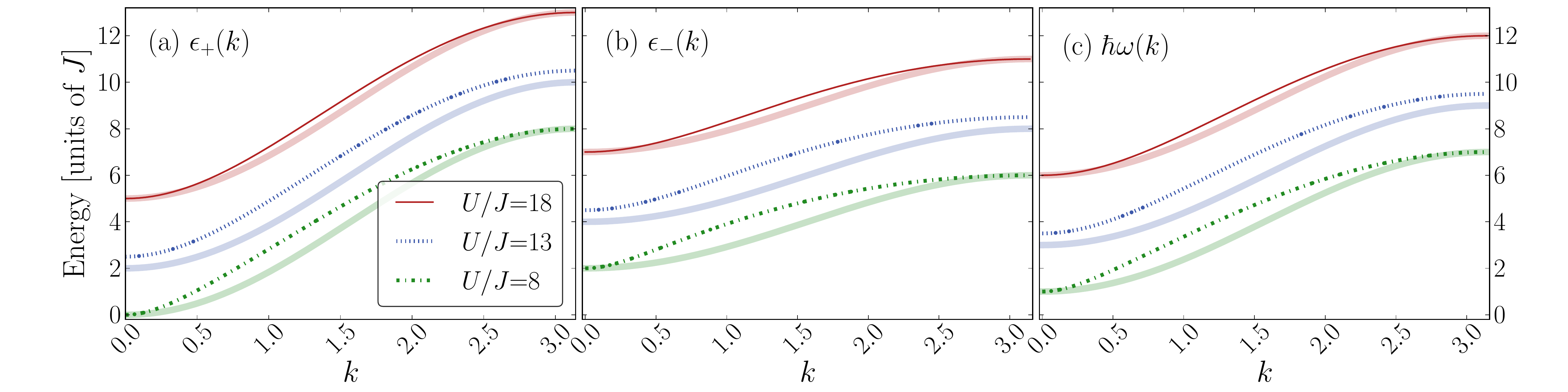}
  \end{center}
  \caption{\label{fig:disp}\figpreamble Quasiparticle dispersions (Eq.~\ref{eq:pmdisp} and
    Eq.~\ref{eq:dispomega}) at $\nnot=1$ for different interaction strengths (thin lines).
    Curvatures can deviate significantly from the cosine form of the strong coupling limit (thick
    lines). The width and the gap of the quasiparticle bands depend on the type of 
    quasiparticle. At $U/J=8$, the gap of the $+$-particle closes, signaling the breakdown of the
    UF approximation.}
\end{figure*}

Here, we follow the standard procedure of Jordan and Wigner \cite{Jordan1928} and introduce
auxiliary fermion operators $c_{j,\sigma}$ with number operators $n_{j,\sigma} =
c^\dagger_{j,\sigma}c_{j,\sigma}$ and anti-commutation relations
\begin{align}
  \anticom{c_{j,\sigma}}{c^\dagger_{j',\sigma'}} & =
  \delta_{j,j'}\delta_{\sigma,\sigma'}\,,\notag\\%
  \anticom{c^\dagger_{j,\sigma}}{c^\dagger_{j',\sigma'}}&= \anticom{ c_{j,\sigma}}{c_{j',\sigma'} }=
  0\,.%
\end{align}
Using non-local string operators,
\begin{align}
  Z_{j,\plus}&=e^{i\pi\sum_{\sigma,j'<j}n_{j',\sigma}}\,,\notag\\
  Z_{j,\minus}&=Z_{j,\plus}e^{i\pi n_{j,\plus}}\label{eq:string}\,,
\end{align}
we  relate the auxiliary bosonic operators to the fermionic ones:
\begin{align}
  b_{j,\sigma}&=Z_{j,\sigma}c_{j,\sigma}\,.
  \label{eq:jwtrans}
\end{align}
The string operator $Z_{j,\sigma}$ counts the parity of the number of fermions accumulated over all
sites $j'<j$ (including the $\plus$-fermion on site $j$ if $\sigma=\minus$) and obeys the relations
\begin{align}
  Z^\dagger_{j,\sigma}=Z_{j,\sigma}\,, \quad Z^2_{j,\sigma}=1\,.
\end{align}
As a consequence, the number operators within both fermionic and bosonic representations coincide:
\begin{align}
  b^\dagger_{j,\sigma}b_{j,\sigma}=c^\dagger_{j,\sigma}c_{j,\sigma}\,,
\end{align}
and the original atom number operator can be written as
\begin{align}
  n_j=n_{j,\plus}-n_{j,\minus}+\nnot\label{eq:fermionstoatoms}\,.
\end{align}

Due to the fermionic statistics, the hard-core conditions (\ref{eq:constraint1}) are satisfied
automatically. The remaining constraint (\ref{eq:constraint2}) can be formally accounted for by the
global projector $\gproj=\prod_j\lproj$, with $\lproj=(1-n_{j,\plus}n_{j,\minus})$ eliminating
states with both species on the same site. It is now possible to show that, within the truncated
Hilbert space, the original Hamiltonian (\ref{eq:bhHamiltonian}) can be exactly represented by the
following fermionic model:
\begin{multline}
\label{eq:hfermionic}
H=\sum_j\gproj \bigg\{ -J(\nnot+1)c_{j,\plus}^\dagger c_{j+1,\plus}\,-\,J\nnot c^\dagger_{j+1,\minus}c_{j,\minus}\\
  -J\sqrt{\nnot(\nnot+1)} \left( c_{j,\plus}^\dagger
    c^\dagger_{j+1,\minus}\,-\,c_{j,\minus}c_{j+1,\plus} \right) +\text{h.c.}\\
  +\frac{U}{2}(n_{j,+}+n_{j,-}) \bigg\} \gproj\,.
\end{multline}
We note that the effective hopping amplitudes for the two different flavors differ by the bosonic
enhancement factor of Eq.~(\ref{eq:sbmapping}).

\subsection{Exact diagonalization within the approximation of unconstrained fermions}
\label{sec:unconstrained}
In practice, it is difficult to take care analytically of the projector $\gproj$. We will thus carry
out the calculations in the approximation of unconstrained fermions (UF), $\gproj\rightarrow 1$,
leading to a quadratic Hamiltonian that can be diagonalized exactly. We will see that the UF
approximation is justified because the main source of creation of double occupancies would be a
local pairing mechanism, which, in the fermionic representation, is suppressed by the statistics of
the auxiliary particles.

The Hamiltonian (\ref{eq:hfermionic}) with $\gproj\equiv1$ can be rewritten in momentum space as
\begin{multline}
  \label{eq:hmom}
  H_{\text{UF}} = \sum_{\sigma,k} E_{\sigma}(\k)c^\dagger_{\k,\sigma}c_{\k,\sigma}\\
  +\sum_{k} \Delta(k)(c^\dagger_{\k,\plus}c^\dagger_{-\k,\minus}\!\!-c_{-\k,\minus}c_{\k,\plus})\,,
\end{multline}
with the bare dispersions
\begin{subequations}
  \label{eq:baredisp}
  \begin{align}
    \label{eq:baeedisp+}
    E_{\plus}(k) & = -2J(\nnot+1)\cos(\k)+U/2\,,\\
    \label{eq:baredisp-}
    E_{\minus}(k) & = -2J\nnot\cos(\k)+U/2\,,
  \end{align}
\end{subequations}
and an antisymmetric pairing parameter
\begin{align}
  \Delta(k)=i\,2J\sqrt{\nnot(\nnot+1)}\sin(\k)\,,
\end{align}
which obeys $\Delta(-\k)=-\Delta(\k)=\Delta^*(\k)$.  In analogy to the Gutzwiller approximation
\cite{Gutzwiller1963, Ogawa1975, Sciro2011}, the accuracy of the UF approximation can be estimated
\emph{via} the translation-invariant expectation value of the local projector
\begin{align}
  \label{eq:guwiapprox}
  p_{\plus\minus}=1-\avg{\mathcal{P}^2_j(t)}=\avg{n_{j,\plus}n_{j,\minus}}\,.
\end{align}
This quantity is a measure for the population of unphysical states and gives the order of magnitude
of the error in local observables (due to translational invariance site indices of observables can be dropped).
Additionally, we will study the quality of the relaxation of the
constraint (\ref{eq:constraint2}) by comparison to the numerically exact DMRG method.

The quadratic Hamiltonian $H_{\text{UF}}$ can be diagonalized via a Bogolyubov transformation by
introducing the quasiparticle operators
\begin{subequations}
  \label{eq:btransf}
  \begin{align}
    \gamma_{\k,\plus}^\dagger & =
    u(\k)c^\dagger_{\k,\plus}+v(\k)c_{-\k,\minus}\,,\label{eq:plusqp}\\
    \gamma_{\k,\minus}^\dagger & = u(\k)c^\dagger_{\k,\minus}-v(\k)c_{-\k,\plus}\,.\label{eq:minusqp}
  \end{align}
\end{subequations}
The functions $u(\k)$ and $v(\k)$ fulfill the relations
\begin{subequations}
  \begin{align}
    u(-\k)=u(\k)=u^*(-\k)\,,\\
    v(-\k)=-v(\k)=v^*(\k)\,,
  \end{align}
\end{subequations}
and are determined by the following expressions:
\begin{subequations}
  \label{eq:btrans}
  \begin{align}
    u(\k)&=\cos\left(\text{atan}\left(\frac{-2i\Delta(\k)}{E_{\plus}(\k)+E_{\minus}(k)}
      \right)/2\right)\label{eq:u}\\
    &= 1+\mathcal{O}\left(\frac{J^2}{U^2}\right)\label{eq:upert} \,,\\
    v(\k)&=i\sin\left(\text{atan}\left(\frac{-2i\Delta(\k)}{E_{\plus}(\k)+E_{\minus}(k)}
      \right)/2\right)\label{eq:v}\\
    &=\,i\frac{2J\sqrt{\nnot(\nnot+1)}}{U}\sin(\k)+\mathcal{O}\left(\frac{J^3}{U^3}
    \right)\label{eq:vpert}\,.
  \end{align}
\end{subequations}
We infer from the above equations that the $\plus$--modes are excess particles each dressed with
absent holes and the $\minus$--modes are holes dressed with absent excess particles. This is
particularly evident from the perturbative expressions (\ref{eq:upert}) and (\ref{eq:vpert}). We
also note that the quasiparticle operators (\ref{eq:btransf}) can be interpreted as Dirac spinors
\cite{Huber2009}.

Using the quasiparticle operators, the Hamiltonian can be written in the diagonal form
\begin{equation}
  H_{\text{UF}} = \sum_{k,\sigma}
  \epsilon_{\sigma}(k)\gamma_{\k,\sigma}^\dagger\gamma_{\k,\sigma}\,.\label{eq:hdiag}
\end{equation}
The dispersion relation of the individual quasiparticles is
\begin{equation}
  \epsilon_\sigma(\k)=-\sigma J\cos(\k)+\hbar\omega(\k)\label{eq:pmdisp}\,.
\end{equation}
Here $2\hbar\omega(\k)$ is the energy of a pair of two distinct types of quasiparticles with
opposite momenta, which is given by
\begin{subequations}
  \begin{align}
    2 \hbar \omega(k) & = \sqrt{\left[E_+(k)+E_-(\k)\right]^2+4|\Delta(\k)|^2}\label{eq:dispomega}
    \\
    & = U-2J(2\nnot+1)\cos(\k)+\mathcal{O}\left(\frac{J^2}{U}\right)\,.\label{eq:disppert}
  \end{align}
\end{subequations}
The exact dispersion relations for different interaction strengths are displayed in
Fig.~\ref{fig:disp}, together with the first order expansion in $J/U$ (strong coupling
expansion). One can see that the profiles rapidly differ from their limiting cosine shape as the
interactions are lowered. Eqs.~\eqref{eq:baredisp} and \eqref{eq:dispomega} show that the width of
the energy bands depends only on the hopping amplitude $J$ and on the average
filling $\bar n$ (\textit{via} the Bose enhancement factor), but does not
depend on the interaction strength. At large interaction strengths, the energy
gap is proportional to the interaction strength and the gap of the
`$\plus$'-quasiparticles is strictly positive when the interaction is above a
certain threshold:
\begin{equation}
  U/J>4(\nnot+1)\,.\label{eq:regime}
\end{equation}
Below this threshold our UF approximation breaks down.  For $\bar
n=1$, the range of validity of our model is thus limited to $U/J > 8$, which is above the
superfluid to Mott-insulator transition $(U/J)_{\text c}\approx 3.4$, but significantly lower than
the mean-field transition $(U/J)_{\text c}\approx 12$. A description of the phase transition might
be achieved by introducing auxiliary operators on the basis of a coherent-state representation (see
e.g. \cite{Huber2007}), but this goes beyond the scope of this work.

The slope of the dispersion relations $\epsilon_{\sigma}(k)$ describes the group velocity of the
quasiparticles. Of particular interest will be the relative velocity of pairs of quasiparticles of
distinct types and opposite momenta:
\begin{align}
  \text{v}(\k)=2\frac{d}{d k}\omega(k)\,,
\end{align}
whose maximal value
\begin{align}
  \text{v}_{\text{max}}=\text{max}_{k}\left|\text{v}(\k)\right|\label{eq:vmax}
\end{align}
plays an important role in the characterization of the non-equilibrium properties. This maximal
velocity corresponds to the point where the curvature of $\omega(k)$ changes sign. It is located at
$|k| \approx \pi/2$ at large interaction strengths and is shifted toward lower momenta at smaller
interaction strengths, as can be seen in Fig~\ref{fig:disp}. In the relevant interaction regime
(\ref{eq:regime}), the maximum velocity is well approximated by:
\begin{align}
  \label{eq:vel}
  \text{v}_{\text{max}}\approx\frac{2J(2\nnot+1)}{\hbar}\left(1-\frac{8\nnot(\nnot+1)J^2}{(2\nnot+1)^2U^2}
  \right)+\mathcal{O}\left(\frac{J^4}{U^3}\right)\,.
\end{align}
In particular, one sees that $\text v_{{\text{max}}}$ is a decreasing function of $U/J$.

For the strictly positive quasiparticle energies (\ref{eq:regime}), the ground state at a value of
$U$ and $J$ is the quasiparticle vacuum
\begin{subequations}
  \label{eq:gswavefunction}
  \begin{align}
    \ket{\psi_0(U/J)}&=\prod_{\k}v^{-1}(k)\gamma_{\k,\plus}\gamma_{-\k,\minus}\fockket \\
    &=\prod_{\k}(u(\k)+v(\k)c^\dagger_{k,\plus} c^\dagger_{-k,-})\fockket\,.
  \end{align}
\end{subequations}
The ground state at infinitely strong interactions, i.e. the Fock state with $\nnot$ particles per
site, $\fockket$, represents the vacuum of the bare excess particles and holes.

\subsection{Local observables and correlation functions}\label{sec:observables}

We summarize in this Sec. some general properties of the correlation functions in the
quasiparticle formalism that will be used later to derive ground state and non-equilibrium
properties of the system.
 
For the ground state (\ref{eq:gswavefunction}), but also for the time-dependent wave functions
(\ref{eq:timeevolution}) which will be introduced in section \ref{sec:timeevolution}, the only
non-vanishing single-particle correlation functions are
\begin{subequations}
  \label{eq:corrft}
  \begin{align}
    g^{\sigma,\sigma}_{d}&=\avg{c^\dagger_{j+d,\sigma}c_{j,\sigma}}\notag\\
    & = \frac{1}{2\pi}\int_{-\pi}^{\pi}\mathrm{d}k\;e^{-ikd}
    \avg{c^\dagger_{k,\sigma}c_{k,\sigma}}\,,\\
    g^{\sigma,\bar{\sigma}}_{d}&= \avg{c_{j+d,\sigma}c_{j,\bar\sigma}}\notag=
    \avg{c^\dagger_{j,\sigma}c^\dagger_{j+d,\bar\sigma}}^{\!*}\\
    &= \frac{1}{2\pi}\int_{-\pi}^{\pi}\mathrm{d} k \;e^{-ikd} \avg{c_{k,\sigma}c_{-k,\bar\sigma}}\,,
    \label{eq:corrftb}
  \end{align}
\end{subequations}
where $\bar{\sigma}=-\sigma$ and the thermodynamic limit has been taken.  Possible time dependence
(equal time) is implicit and expectation values are site-independent for the homogeneous systems
under consideration.  We note that the correlations of the different types are equivalent:
\begin{equation}
  g^{\plus,\plus}_{d} = g^{\minus,\minus}_{d}\,, \quad g^{\plus,\minus}_{d} = g^{\minus,\plus}_{d}\,.
\end{equation}
Therefore, also the quasiparticle densities do not depend on the flavor and we can define a single
density of excitations:
\begin{align}\label{eq:nex}
  \nex=\avg{n_{j,+}}+\avg{n_{j,-}}=2g^{\sigma,\sigma}_{0}\,.
\end{align}
Since the Hamiltonian is quadratic, correlations of the occupancies can be related to the
single-particle correlations using Wick's theorem, which gives us
\begin{align}
  G^{\sigma,\sigma'}_{d} & = \avg{n_{j+d,\sigma}n_{j,\sigma'}}-\avg{n_{j+d,\sigma}}
  \avg{n_{j,\sigma'}}\\
  & = -\sigma\sigma'|g^{\sigma,\sigma'}_{d}|^2\,.\label{eq:oocorr}
\end{align}
In the special case $d=0$, the fermionic statistics, together with the symmetry with respect to 
exchange of fermionic flavor, imply that the on-site correlator
$G^{\sigma,\bar\sigma}_{d=0}=|g^{\sigma,\bar\sigma}_{d=0}|^2$ vanishes and the local double
occupancy factorizes:
\begin{align}
  \avg{n_{j,\sigma}n_{j,\bar\sigma}}=\avg{n_{j,\sigma}}\avg{n_{j,\bar\sigma}}
  =\nex^2/4\,. \label{eq:occfactorization}
\end{align}
The density of excitations thus fully determines all local properties, including the atom
number fluctuations:
\begin{equation}
  \label{eq:atnumfluc}
  f=\avg{(n_{j}-\nnot)^2}=\nex(1-\nex/2)\,.
\end{equation}

Atom correlations can be related to correlations of auxiliary particles. Making use of
Eq.~(\ref{eq:fermionstoatoms}), we can for example express atomic density correlations in the following
way:
\begin{align}
  C_d & = \avg{n_jn_{j+d}}-\avg{n_j}\avg{n_{j+d}}\label{eq:nncorrdef}\\
  & = \sum_\sigma \left( G^{\sigma,\sigma}_{d}-G^{\sigma,\bar\sigma}_{d}\right)\\
  & = -2(|g^{+,+}_{d}|^2 +|g^{+,-}_{d}|^2)\,.\label{eq:nncorr}
\end{align}
Ultracold atom experiments with single-site resolved imaging
\cite{Bakr2009, Sherson2010} can access the parity $s_j=e^{i\pi(n_j-\nnot)}$ rather than the density itself. The expression for
parity correlations turns out to be similar to that of density correlations:
\begin{align}
  S_d & = \avg{s_js_{j+d}}-\avg{s_j}\avg{s_{j+d}}\\
  & = 4\sum_\sigma \left(G^{\sigma,\sigma}_{d}+G^{\sigma,\bar\sigma}_{d}\right)\,.
\end{align}
Both density and parity correlations are particularly simple to evaluate within the present
approach. Correlations including the non-local string operator (\ref{eq:string}), such as the
single-particle correlations $\avg{a^\dagger_ja_{j+d}}$, can also be computed, but they require the
evaluation of the Toeplitz determinant \cite{Barouch1971b}. Interestingly, the fermionic string
(\ref{eq:string}) is equivalent to the string operator recently measured by
Endres \etal \cite{Endres2011}.

\begin{figure*}[ht]
  \begin{center}
    \includegraphics[width=0.99\textwidth]{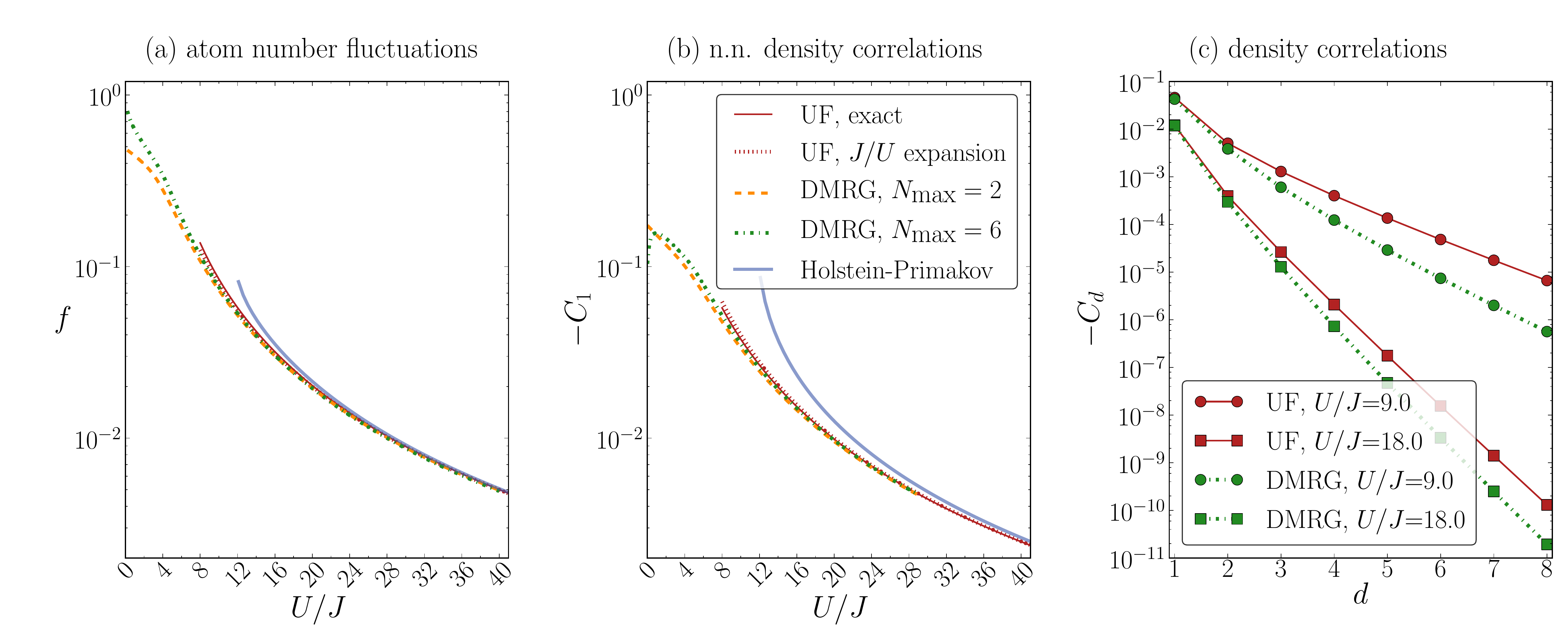}
  \end{center}
  \caption{\label{fig:gscorr} \figpreamble Ground state properties at $\nnot=1$. The
    numerical evaluation of the UF equations is compared with the strong coupling expansion, as well
    as with exact DMRG simulations of the Bose--Hubbard model with the local Hilbert space truncated
    at a maximum site occupancy of either $N_{\text{max}}=2$ or $N_{\text{max}}=6$. (a) Atom number
    fluctuation $f$ as a function of the final interaction strength $U/J$. (b) Nearest-neighbor
    density correlation $C_{d=1}$ as a function of the final interaction strength $U/J$. (c) Density
    correlations as a function of the distance $d$.}
\end{figure*}

\section{Equilibrium properties of the Mott-insulating phase}
\label{sec:equ_prop}
In this section we discuss the equilibrium properties of the Mott-insulating phase derived within
the unconstrained fermion approximation.

As argued in the preceding section, the observables are related to single-particle correlations
(\ref{eq:corrft}), which for the ground state (\ref{eq:gswavefunction}) can be evaluated
straightforwardly:
\begin{align}
  \avg{c_{k,\sigma}^\dagger c_{k',\sigma}}&=-\delta_{k,k'}v^2(\k)\,,
  \label{eq:equalcorr}\\
  \avg{c_{k,\sigma}c_{-k',\bar\sigma}}&=\delta_{k,k'}u(\k)v(\k)\,,
  \label{eq:diffcorr}
\end{align}
with the coefficients $u(\k)$ and $v(\k)$ given in Eq. (\ref{eq:btrans}). The local density of
excitations (\ref{eq:nex}) can thus be calculated from
\begin{align}
  \nex = -\frac{1}{\pi}\int_{-\pi}^{\pi}\mathrm{d}k\;v^2(\k)\,.\label{eq:gsnex}
\end{align}
In the case of strong interactions, one can also derive an explicit expression from the expansion
(\ref{eq:upert}) and (\ref{eq:vpert}) of the coefficients $u(k)$ and $v(k)$:
\begin{align}
  \nex=\frac{2J^2}{U^2}\nnot(\nnot+1)+\mathcal{O}\left(\frac{J^4}{U^4}\right).
  \label{eq:neq}
\end{align}
Combining Eqs.~\eqref{eq:occfactorization} and \eqref{eq:neq} gives an estimate of the occupation
of unphysical states (\ref{eq:guwiapprox}),
\begin{align}
  p_{\plus\minus}=\nex^2/4 \lesssim \left(8\left(1+1/\nnot\right)\right)^{-2}\,,
  \label{eq:guwiapprox2}
\end{align}
where the right-hand side stems from the evaluation of (\ref{eq:neq}) at the lowest interaction
considered (\ref{eq:regime}). For $\nnot=1$, $p_{\plus\minus}$ is less than $6\%$ and we expect the
error on local expectation values to be of similar magnitude. With this at hand, we can now consider
the behavior of the density correlations in the ground state. In Fig.~\ref{fig:gscorr}(a), the
atom number fluctuation $f=\nex(1-\nex/2)$ is evaluated numerically using (\ref{eq:gsnex}) and
compared to the strong coupling expansion (see also \cite{Freericks1994})
\begin{equation}
  f=\frac{2J^2}{U^2}\nnot(\nnot+1)+\mathcal{O}\left(\frac{J^4}{U^4}\right)\,,
\end{equation}
as well as to the results obtained from DMRG simulations with a truncation of the site occupancy to
$N_{\text{max}}=2$ or $N_{\text{max}}=6$ (system size is 256 sites, 400 DMRG-states are
retained). The predictions of the UF approximation, both from the numerical integration of
(\ref{eq:gsnex}) and from the strong coupling expansion, are in excellent agreement with the DMRG
simulations for all interaction strengths satisfying (\ref{eq:regime}). The accuracy of the truncation
of the local Hilbert space to three states only is also confirmed by the DMRG simulations. Higher
occupancies start to be important only for interaction strengths below the point where the UF
approximation breaks down.

We further compare our results with the ones derived within a Holstein--Primakov approximation of the
slave-boson representation used, e.g., by Huber \etal~\cite{Huber2007}. This approach is equivalent
to the auxiliary boson representation (\ref{eq:sbmapping}) when fully relaxing the constraints
(\ref{eq:constraint1},\ref{eq:constraint2}). We find that the local observables cannot be well
described at intermediate interaction strengths, even though the density of excitations is small
(Fig.~\ref{fig:gscorr}). A similar instability has been observed with slave bosons in
Ref.~\cite{Lu2006}. We will analyze the slave-boson approach in more detail later in the context of
the non-equilibrium dynamics (Sec.~\ref{sec:holstein-primakov}).

We can also evaluate analytically non-local density correlations to second order in $J/U$:
\begin{align}
  C_d=-\frac{J^2}{U^2}\nnot(\nnot+1)\delta_{d,1}
  +\mathcal{O}\left(\frac{J^4}{U^4}\right).\label{eq:gscorrpert}
\end{align}
As shown in Fig.~\ref{fig:gscorr}(b), the above expression only slightly overestimates the amplitude
of the correlations compared to the full numerical evaluation of the integrals (\ref{eq:corrft})
and (\ref{eq:nncorr}) with (\ref{eq:diffcorr}).  We therefore conclude that local observables and
nearest-neighbor correlations in the Mott-insulator regime are well described by a perturbation
expansion to order $J^2/U^2$.  This is no longer the case for longer-range correlations, which are
simply vanishing according to the expansion to second order, whereas the exact DMRG predicts that
they should be finite and exponentially decaying. As can be seen in Fig. \ref{fig:gscorr}(c), the
numerical evaluation of the UF equations provides a much better agreement with the DMRG results. The
correlations at $d=2$ can be almost perfectly reproduced and a similar decay length is found. The
amplitude of the correlations for $d>2$ is overestimated, however, and the discrepancy becomes worse
as $d$ increases.

\section{Quench dynamics -- general description}
\label{sec:timeevolution}
We analyze the quench dynamics of a system prepared initially in a deep Mott-insulating state. We
first derive the general results for the time evolution of the wave function and the correlation
functions and then give explicit expressions for the case where the initial state is a Fock state
(infinite interactions). These results form the basis for the detailed discussion of the physical
properties of the quench dynamics in the subsequent Sec. \ref{sec:prop-quas}.

\subsection{Time-dependent wave function and correlations}
The initial state considered is the ground state at some values of $J$ and $U$ satisfying the
condition (\ref{eq:regime}) and takes the form
\begin{align}
  \ketinit=\prod_{\k} \left( u_0(\k)+v_0(\k)c^\dagger_{k,\plus} c^\dagger_{-k,-} \right)\fockket\,.\label{eq:initialstate}
\end{align}
The time-evolution of this state under the Hamiltonian $H_{\text{UF}}$ reads
\begin{align}
  \label{eq:timeevolution}
  \ket{\psi(t)}&=e^{-iH_{\text{UF}}t/\hbar}\ket{\psi_{\text{\tiny init}}}\notag\\
  &=\prod_{\k}\left(\bar{u}(\k)-\bar{v}(\k)e^{-i2\omega(k)t}
    \gamma_{\k,\plus}^\dagger\gamma^\dagger_{-\k,\minus}\right)\ket{\psi_0(U/J)} \,,
\end{align}
with
\begin{subequations}
  \begin{align}
    \bar{u}(k)={u}(k){u_0}(k)-{v}(k){v_0}(k)\,,\\
    \bar{v}(k)={v}(k){u_0}(k)-{u}(k){v_0}(k)\,.
  \end{align}
\end{subequations}
For the wave function (\ref{eq:timeevolution}), the non-vanishing equal-time single-particle
correlations evaluate to
\begin{subequations}
  \label{eq:corrk}
  \begin{gather}
    \begin{split}
      \av{c_{k,\sigma }(t)c_{-\k',\bar\sigma}(t)} = \delta_{k,k'}u(k)\bar{u}(k) \left[
      e^{-2i\omega(k)t} u(k) \bar{v}(k) - \bar{u}(k){v}(k) \right] \\
      +\delta_{k,k'}v(k)\bar{v}(k) \left[e^{2i\omega(k)t} v(k) \bar{u}(k) - \bar{v}(k){u}(k) \right]
    \end{split} \label{eq:corrk1}
    \intertext{and}
    \begin{multlined}[b]
     \av{c^\dagger_{k,\sigma }(t)c_{\k',\sigma}(t)} = \delta_{k,k'} \left[2\cos(2\omega
     (k)t)u(k)\bar{u}(k)v(k)\bar{v}(k) \right.\\
     \left. u^2(k)\bar{v}^2(k)-v^2(k)\bar{u}^2(k) \right]\,.
    \end{multlined} \label{eq:corrk2}
  \end{gather}
\end{subequations}
Based on these expressions, the expectation values of any observables be either calculated
analytically, when the strong coupling expansion holds, or computed numerically for lower
interactions (see \ref{sec:observables}).

\subsection{Strong coupling expansion}
\label{sec:pertexp}
For concreteness, we focus now on a quantum quench starting from the Fock state with filling
$\bar{n}$ by setting $u_0(\k)=1$, $v_0(\k)=0$ and thus $\bar{u}(k)=u(k)$, $\bar{v}(k)=v(k)$. In this
case, the expansion in $J/U$ leads to
\begin{subequations}
  \label{eq:corrkpert1}
  \begin{gather}
    \begin{split}
      \av{c_{k,\sigma }(t)c_{-\k,\bar\sigma}(t)} = i\frac{2J\sqrt{\nnot(\nnot+1)}}{U}\sin(\k)
      \left[ e^{-2i\omega(k) t} - 1 \right]\\
      +\mathcal{O}\left(\frac{J^2}{U^2}\right)\,,
    \end{split}\label{eq:corrkpert1}\\
    \begin{split}
      \av{c^\dagger_{k,\sigma }(t)c_{\k,\sigma}(t)} = \frac{8J^2\nnot(\nnot+1)}{U^2}\sin^2(\k)
      \left[\cos(2\omega(k) t) - 1\right] \\
      +\mathcal{O}\left(\frac{J^4}{U^4}\right)\,,
    \end{split}\label{eq:corrkpert2}
  \end{gather}
\end{subequations}
where $\omega(k)$ stands for the dispersion in the strong coupling expansion (\ref{eq:disppert}).

Within this expansion, the dynamics of the local density of excitations can be expressed
in terms of the Bessel functions of the first kind, $\Jb_n(z)=\frac{i^{-n}}{2\pi}\int_{-\pi}^{\pi}d k
e^{-i z\cos(k)+n k}$. One gets
\begin{equation}
  \nex(t)\approx\frac{8\nnot(\nnot+1)J^2}{U^2}\left[1-\cos(Ut/\hbar) \left(\Jb_{2}(\vinfty
      t)+\Jb_{0}(\vinfty t)\right)\right]\label{eq:approxnex}\,,
\end{equation}
with $\vinfty=2J(2\nnot+1)/\hbar$. In the relevant interaction regime, the population of unphysical
states $p_{\plus\minus}(t)$ thus remains as small as in the ground state (\ref{eq:guwiapprox2}) and
we expect the UF approximation to be well behaved in general. It is, however, important to note, that
the expansion is not rigorous since the approximate dispersion (\ref{eq:disppert}) is multiplied
by time, which is unbounded.

The single-particle correlators required to derive non-local density correlations (\ref{eq:nncorr}) read
  \begin{align}
    g_{\ell}^{\sigma,\bar\sigma}(t) & \approx i\frac{2J\sqrt{\nnot(\nnot+1)}}{2\pi U}
    \int d k\; e^{i\k\ell}\sin(\k) \left[e^{-2i\omega(k) t}-1\right] \notag\\
    & = 
    \begin{multlined}[t][0.825\columnwidth]
      (-i)^{d+1}\frac{\sqrt{\nnot(\nnot+1)}J}{U} \left[e^{iUt/\hbar} \big(\Jb_{d+1}(\vinfty t) \right.\\
      \left.+\Jb_{d-1}(\vinfty t)\big)+\delta_{d,1}\right] \;.
    \end{multlined} \label{eq:sgpartcorr}
  \end{align}
Making use of the identity
\begin{equation}
  \label{eq:besselidentity}
  \Jb_{d+1}(z)+\Jb_{d-1}(z)=\frac{2d}{z}\Jb_{d}(z) \;,
\end{equation}
we obtain the following expressions for the non-local density correlations:
\begin{subequations}
  \label{eq:corrapprox}
  \begin{align}
    \label{eq:corrapprox1}
    C_{d=1}(t) &\approx -\left(\frac{2\nnot(\nnot+1)Jd}{U}\right)^2
    \left(\frac{\Jb_d(\vinfty t)}{\vinfty t}2\cos(Ut/\hbar)+1\right) ,\\
    C_{d>1}(t) &\approx -\left(\frac{\nnot(\nnot+1)Jd}{U}\right)^2 \left( \frac{\Jb_d(\vinfty
        t)}{\vinfty t}\right)^2 .
    \label{eq:corrapprox2}
  \end{align}

\end{subequations}
We note that the interaction strength $U$ is involved only in the magnitude of the correlations for
$d>1$, via the dimensionless parameter $J/U$. In the case of nearest-neighbor correlations, we find an
additional oscillation of the amplitude with frequency $U/h$.

\subsection{Accuracy of the UF approximation}
\label{sec:accuracy}
\begin{figure*}[ht]
  \begin{center}
    \includegraphics[width=\textwidth]{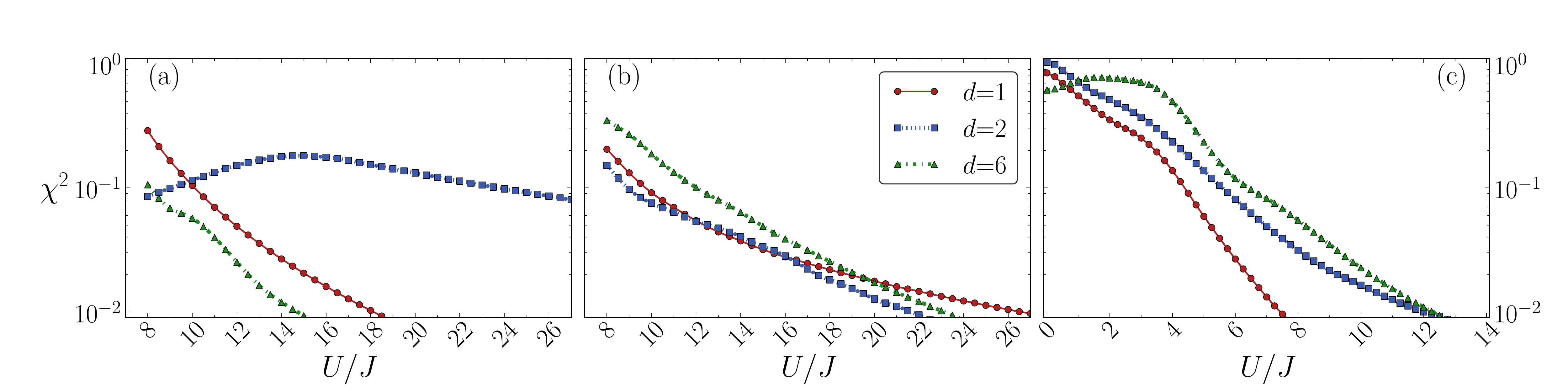}
  \end{center}
  \caption{\label{fig:diff}\figpreamble Root-mean-square differences $\chi^2$ of the density correlations
    (\ref{eq:chi2}) obtained from different degrees of approximation.  (a) The strong coupling
    expansion is compared to the numerical evaluation of the UF approximation. (b) The numerical
    evaluation of the UF approximation is compared to the exact DMRG simulation of the Bose--Hubbard
    model with a local site occupancy truncated at $N_{\text{max}}=2$. (c) The DMRG simulation with
    $N_{\text{max}}=2$ is compared to the DMRG simulation truncated at $N_{\text{max}}=6$.}
\end{figure*}
In this section, we analyze the accuracy of the successive approximations that lead from the
Bose--Hubbard model to the UF approximation and its strong coupling expansion. For this purpose, we
introduce the root-mean-square differences
\begin{equation}
  \label{eq:chi2}
  \chi_{d}^2=\frac{\int_0^{t_{\text{max}}} dt \left(C_d^{(2)}(t)-C_d^{(1)}(t)\right)^2}
  {\int_0^{t_{\text{max}}} dt \left(C_d^{(2)}(t)\right)^2}\,,
\end{equation}
where $C_{d}^{(1)}$ and $C_{d}^{(2)}$ are the density correlations predicted using two different
level of approximations.  By observing the dependency of $\chi^{2}_{d}$ with the distance $d$, we can
verify, in particular, whether a given approximation breaks down at large times. For non-averaged results we refer to Sec. \ref{sec:prop-quas}. We use
$t_{\text{max}}=3\hbar/J$, the maximal time accessible by our DMRG simulations.  We use a DMRG
algorithm in the thermodynamic limit \cite{Vidal2007,Mcculloch2008}, with a second-order
Suzuki--Trotter decomposition of time step $\Delta t=0.02\hbar/U$, and we retain 2400 states.
The numerical error is always smaller than the symbol size and line width.

In Fig.~\ref{fig:diff}(a), we first compare the strong coupling expansion (\ref{eq:corrapprox}) and
the numerical evaluation of the UF approximation. We find that the expansion is relatively accurate
($\chi_{d}^{2}<10^{-1}$) down to interactions $U/J\sim10$, except for the $d=2$ correlation, which
only slowly converges to the exact results when $U/J$ is increased. We recall here that a similar
accuracy is reached for the ground state correlations (Fig.~\ref{fig:gscorr}).

In Fig.~\ref{fig:diff}(b), we then compare the numerical evaluation of the UF approximation with the
exact DMRG simulation of the Hamiltonian (\ref{eq:hfermionic}), i.e. the Bose--Hubbard model when
the site occupancy is truncated to $N_{\text{max}}=2$. The UF approximation appears to be accurate
within $\chi_{d}^{2}<10^{-1}$ for $U/J\gtrsim12$. As we will show in Sec. \ref{sec:corrpropagation},
the UF approximation still qualitatively describes the dynamics between $U/J\approx12$ and $U/J=8$,
which marks the break down of the quasiparticle picture.

Finally, we compare in Fig.~\ref{fig:diff}(c) the predictions of the DMRG simulation when the local
Hilbert space is truncated to a maximum site occupancy $N_{\text{max}}=2$, corresponding to the
model (\ref{eq:hfermionic}), or $N_{\text{max}}=6$. We observe that the error due to the truncation
starts to be significant ($\chi_{d}^{2}>10^{-1}$) only for $U/J<6$, i.e. when the interaction energy
becomes larger than the width of the quasiparticle band.

\subsection{Comparison with the Holstein--Primakov approximation for auxiliary bosons}
\label{sec:holstein-primakov}

In order to generalize the description to higher dimensions, it may appear tempting to fully relax
the hardcore constraint \eqref{eq:constraint1} and work with bosons instead of fermions. The
resulting Hamiltonian is then equivalent to the one derived by Huber \etal~\cite{Huber2007} using
Holstein--Primakov bosons (HP). The resulting equations for the quasiparticles and their dispersions are very similar to those derived in the fermionic
model, except that the coefficient $v(k)$ becomes symmetric instead of antisymmetric. As a
consequence, local pairing of different species is no longer suppressed and the occupation of the
unphysical states becomes much larger than in the fermionic approach. In order to quantify this
effect, one first has to calculate the single-particle correlations. To lowest order in $J/U$, 
the quasiparticle coefficient takes the form $v(k)\stackrel{\text{HP}}{\approx}\frac{2J\sqrt{\nnot(\nnot+1)}}{U}\cos(k)$ and one finds
\begin{equation}
	\label{eq:hpg}
  g_{\ell}^{\sigma,\bar\sigma}(t) \stackrel{\text{HP}}{\approx} \frac{2J\sqrt{\nnot(\nnot+1)}}{2\pi
    U} \int d k \, e^{i\k\ell}\cos(\k) \left[ e^{i\vinfty\cos(k) t}-1 \right]\,,
\end{equation}
which can be compared to the fermionic version (\ref{eq:sgpartcorr}). In particular, one finds that 
the integral \eqref{eq:hpg} has a finite value at $d=0$ and the overcompleteness
\begin{equation}
  p_{\plus\minus}\stackrel{\text{HP}}{\approx}\frac{2\nnot(\nnot+1)J^2}{U^2}\Jb_1(\vinfty t)
\end{equation}
becomes of the same order as the density of excitations and the density correlations. 
This means that physical and unphysical states play an equally important role in the
HP approximation and this approach fails to describe the quench dynamics even in the limit 
$U \gg J$, where the density of excitations is low.
This can be observed for example in the density correlations, which now read
\begin{equation}
	\label{eq:hpcorr}
  C_{d>1}(t) \stackrel{\text{HP}}{\approx} -\left(\frac{\nnot(\nnot+1)J}{2U}\right)^2
  \left[\Jb_{d-1}(\vinfty t)-\Jb_{d+1}(\vinfty t) \right]^{2}.
\end{equation}
The change of sign between the two Bessel functions compared to the UF expressions has a dramatic
effect, since the function in square brackets is now proportional to the derivative of a Bessel function,
instead of a Bessel function itself \eqref{eq:besselidentity}. This leads in particular to the
smearing out of one of the main features of the quench dynamics, namely the propagating correlation
peak that we will describe in Sec.~\ref{sec:corrpropagation}.

\subsection{Limit of non-interacting bosons}
In this section, we complement the preceding analysis of quenches on the strongly interacting side
by the extreme situation of a quench from infinite to zero interactions \cite{Flesch2008,
  Natu2012}. At $U/J=0$, the time evolution is readily described in the Heisenberg picture,
$$\hat{a}_j(t)=\sum_{j'=1}^L V_{j,j'}(t) \hat{a}_{j'}\,,$$
in which individual bosons propagate with free dispersion. In the thermodynamic limit, this yields the
propagator
\begin{align}
  V_{j,j+d}(t)&=\frac{1}{2\pi}\int_{-\pi}^{\pi}dk\exp\left[
    -i\left(2J\cos(k)t/\hbar - kd\right)\right]\nonumber \\
  &=(i)^{d}\Jb_{d}\left(\frac{2Jt}{\hbar}\right)\, .\label{eq:VBessel}
\end{align}
Using the following relation for the initial Fock state:
\begin{align}
  \bra{\nnot} \hat{a}_p^{\dagger}\hat{a}_q\hat{a}_r^{\dagger}\hat{a}_s \ket{\nnot} =
  \nnot^2\delta_{p,q} \delta_{r,s} + \nnot(\nnot+1) (1-\delta_{p,q})\delta_{p,s}\delta_{q,r}
  \,. \nonumber
\end{align}
we can derive explicit equations for the density correlations:
\begin{align}
  \notag
  C_d(t)  & =
  \begin{multlined}[t]
    \nnot^2\Big( \sum_j \Jb_j^2(2Jt/\hbar) \Big)^2 \\
    + \nnot(\nnot+1) \Big( \sum_j \Jb_{j+d}(2Jt/\hbar) \Jb_{j}(2Jt/\hbar) \Big)^2 \\
    - \nnot(\nnot+1) \sum_j \Jb_{j+d}^2(2Jt/\hbar) \Jb_{j}^2(2Jt/\hbar) -\nnot^2
    \end{multlined} \\
    & = - \nnot(\nnot+1) \sum_j \Jb_{j+d}^2(2Jt/\hbar) \Jb_{j}^2(2Jt/\hbar) \,.
  \label{eq:corrnonint}
\end{align}
Here we have used the properties $\Jb_n(u\pm v) = \sum_{m=-\infty}^{\infty} \Jb_{n\mp m}(u)\Jb_m(v)$
and $\Jb_d(0)=0$ for $d\ne 0$.

\section{How quasiparticle pairs carry density correlations across the system}
\label{sec:prop-quas}
We now analyze in detail how correlations spread in the quench dynamics starting from the Fock state
with $\bar n$ atoms per site within the Bose--Hubbard Hamiltonian
(\ref{eq:bhHamiltonian}). The description in terms of fermionic quasiparticles for
intermediate and strong interactions (\ref{eq:timeevolution}), as well as the non-interacting solution
(\ref{eq:corrnonint}), provide a firm basis for the interpretation of the outcome of the DMRG
simulations and of recent experimental results \cite{Cheneau2012} and allow for their extrapolation
at long times, where no analytical solution is available so far.

\subsection{Quasiparticle pairs}
For concreteness, we restrict our discussion in the following to the filling $\nnot=1$, where the
$\plus$-quasiparticles \eqref{eq:plusqp} correspond to doublons and $\minus$-quasiparticles
\eqref{eq:minusqp} to holons. The relevant processes involved in the quench dynamics can be best
understood in the expansion of the wave function \eqref{eq:timeevolution} to lowest order in the
auxiliary fermion operators:
\begin{subequations}
  \label{eq:psipert}
  \begin{align}
    \ket{\psi(t)} & =\ket{\nnot}+i\frac{2\sqrt{2}J}{U}\sum_k
    \sin(k)c_{\k,\plus}^\dagger c_{-\k,\minus}^\dagger\ket{\nnot} \label{eq:psipert1} \\
    & \quad -i\frac{2\sqrt{2}J}{U} \sum_k \sin(k) e^{i6J\cos(k)t/\hbar} c_{\k,\plus}^\dagger
    c_{-\k,\minus}^\dagger \ket{\nnot} \label{eq:psipert2}\,.
  \end{align}
\end{subequations}
In this representation, the state decomposes into two parts: a time-independent part
(\ref{eq:psipert1}) consisting of the Fock state and the symmetric superposition of bound
nearest-neighbor doublon-holon pairs, and a time-dependent part (\ref{eq:psipert2}) describing the
superposition of propagating doublon-holon pairs. The dynamics following the quench is driven by the
propagating pairs, whereas the steady state is solely determined by the bound pairs, as the
contribution of the propagating pairs phases out at long times. At lowest order in $J/U$, the steady
state simply corresponds to the ground state at the final interaction strength. Higher order terms
in the strong coupling expansion would describe the population in the excited states. At $t=0$, the
bound and propagating pairs interfere destructively and one recovers the initial Fock state
$\ket{\nnot}$. Finally, we note that the wave function (\ref{eq:psipert}) is equivalent to the one
obtained within a time-dependent perturbation theory in Appendix \ref{sec:perturbation} and can be
used to derive the perturbative results presented in Sec.~\ref{sec:pertexp}.

The doublon and the holon forming a propagating pair are produced initially on neighboring sites by
a single hopping event and then move in opposite directions. The two quasiparticles are entangled,
since the pair is described by a superposition state:
  $$\left(c_{\k,\plus}^\dagger c_{-\k,\minus}^\dagger - c_{-\k,\plus}^\dagger
    c_{\k,\minus}^\dagger\right) \ket{\nnot}.$$ This
  ensures a constant atomic density and leads to strong bipartite entanglement as the pairs are
  stretched across the system \cite{Calabrese2006}. The momentum distribution of the quasiparticle
  pairs is sine shaped, from which follows that the quasiparticles propagate as a wave packet. The
  maximal weight of the momentum distribution is located at the wave vector $|k|=\pi/2$, where the
  dispersion relation \eqref{eq:disppert} is close to being linear, and is characterized by the
  maximal group velocity $\text{v}_{\text{max}}=6J/\hbar$. The wave-packet structure of the
  propagation can also be made explicit by turning the sum over the momenta in \eqref{eq:psipert2}
  into a sum over the lattice sites:
  \begin{multline}
    \sum_k \sin(k) e^{i6J\cos(k)t/\hbar}c_{\k,\plus}^\dagger
    c_{-\k,\minus}^\dagger\ket{\nnot} \\
    = \sum_{j,d}\frac{(-i)^d\hbar d}{3Jt} \Jb_d(6Jt/\hbar)c_{j,\plus}^\dagger
    c_{j+d,\minus}^\dagger\ket{\nnot}\,.
  \end{multline}
  In the above expression, one immediately recognizes the propagation velocity $6J/\hbar$ in the
  argument of the Bessel functions. However, we expect a large dispersion of the wave packet due to the width
  of the momentum distribution. A detailed description of the propagation of the quasiparticle pairs
  is left for Sec.~\ref{sec:corrpropagation}.

  The situation is somewhat different for weakly interacting bosons. In the non-interacting solution
  \eqref{eq:corrnonint}, the correlation functions result from the interference between free bosons
  propagating with a relative velocity $4J/\hbar$.  Unlike the auxiliary particles in the strongly
  interacting case, the number of free bosons per site is not limited, which leads to some
  qualitative differences that we will discuss later.

  \subsection{Correlation signal in the density correlations}

\begin{figure}[ht]
    \begin{center}
      \includegraphics[width=0.49\textwidth]{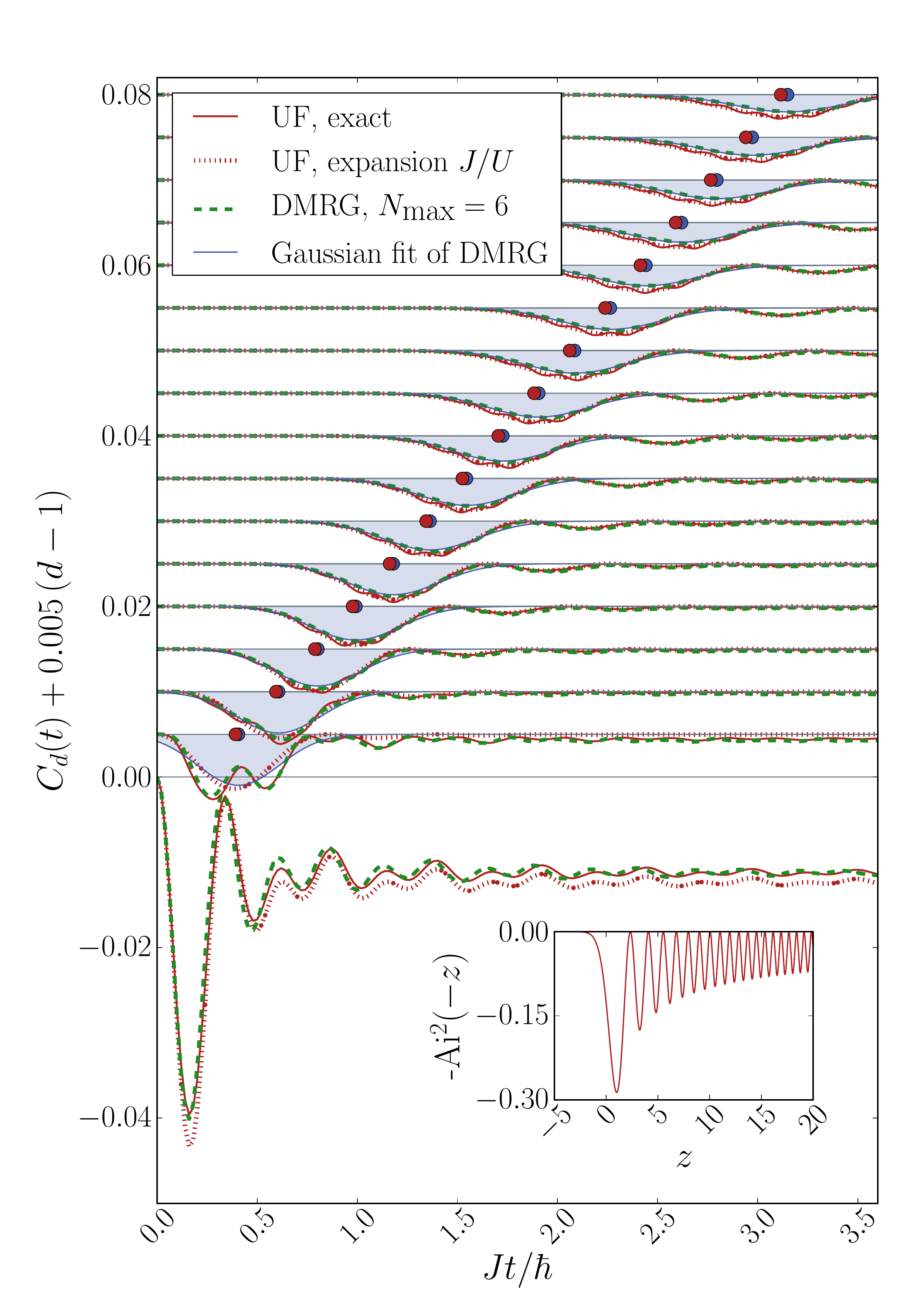}
    \end{center}
    \caption{\label{fig:quenchcorr1}\figpreamble Dynamics of density correlations at distance $d\geq1$ after a
      quench from the Fock state $\ket{\nnot}$ at infinite interactions to a final interaction $U/J=18$. The results
      for different distances $d$ are shifted for clarity by $0.005(d-1)$. We display the results
      obtained from the numerical evaluation of the UF equations, from their strong coupling
      expansion and from exact DMRG simulations. The shaded blue profiles figure a Gaussian fit of
      the correlation signal from the DMRG simulation, after the high-frequency oscillations have been
      filtered out. The filled blue circles mark the center of the fitted profile, i.e. the position
      of the signal. The filled red circles mark the position of the correlation signal
      obtained in the same way from the numerical evaluation of the UF approximation (fit not
      shown). The Airy function that appears in the analytical formulas derived from the UF
      approximation is plotted in the inset.}
  \end{figure}

  The equal-time density correlations $C_d(t)$ in the strongly interacting limit exhibit a very
  peculiar feature for $d\geq2$, namely, the presence of a negative signal, a dip, propagating to
  larger distances at longer times. This can be seen, for example, in Fig.~\ref{fig:quenchcorr1},
  where we display the time evolution of these correlations for $U/J=18$, as predicted by the UF
  approximation and the DMRG simulation (which are in remarkable agreement). This characteristic
  signal is already present in the perturbative result and can be attributed to the propagating
  quasiparticle pairs, illustrating the interest of this picture. 

The structure of the
  nearest-neighbor correlation is more complicated. In the long-time limit and within the strong coupling
  expansion, the nearest-neighbor correlation reaches the value corresponding to the ground state at
  the final interaction strength. At short time, it exhibits oscillations driven by the interaction
  strength $U$ and corresponding to the interaction of a holon (doublon) of the bound pair
  \eqref{eq:psipert1} with a doublon (holon) of a propagating pair \eqref{eq:psipert2}. In the first
  order of the strong coupling expansion (\ref{eq:psipert}), the bound pairs extend only over a
  distance $d=1$ but in the full numerical integration they can spread over larger distances
  [cf.~Fig.~\ref{fig:gscorr}(c)], leading to additional oscillations for $d=2$. These oscillations
  are clearly visible in numerical evaluations of the correlations within the UF approximation, as
  well as in the DMRG simulations (Fig.~\ref{fig:quenchcorr1}).

  While the UF approximation is almost exact at short distances, it overestimates the weak
  oscillations with period $h/U$ at larger distances. We found that these stem from terms of order
  $J^4/U^4$ dominating the doublon-doublon and holon-holon correlations. These
  oscillations are also present in the DMRG simulations, but with a much lower amplitude.
  Interestingly, these terms cancel in the parity correlations studied in \cite{Cheneau2012}, where
  the UF approximation is in even better agreement with the exact simulations.

\begin{figure}[ht]
    \begin{center}
      \includegraphics[width=0.49\textwidth]{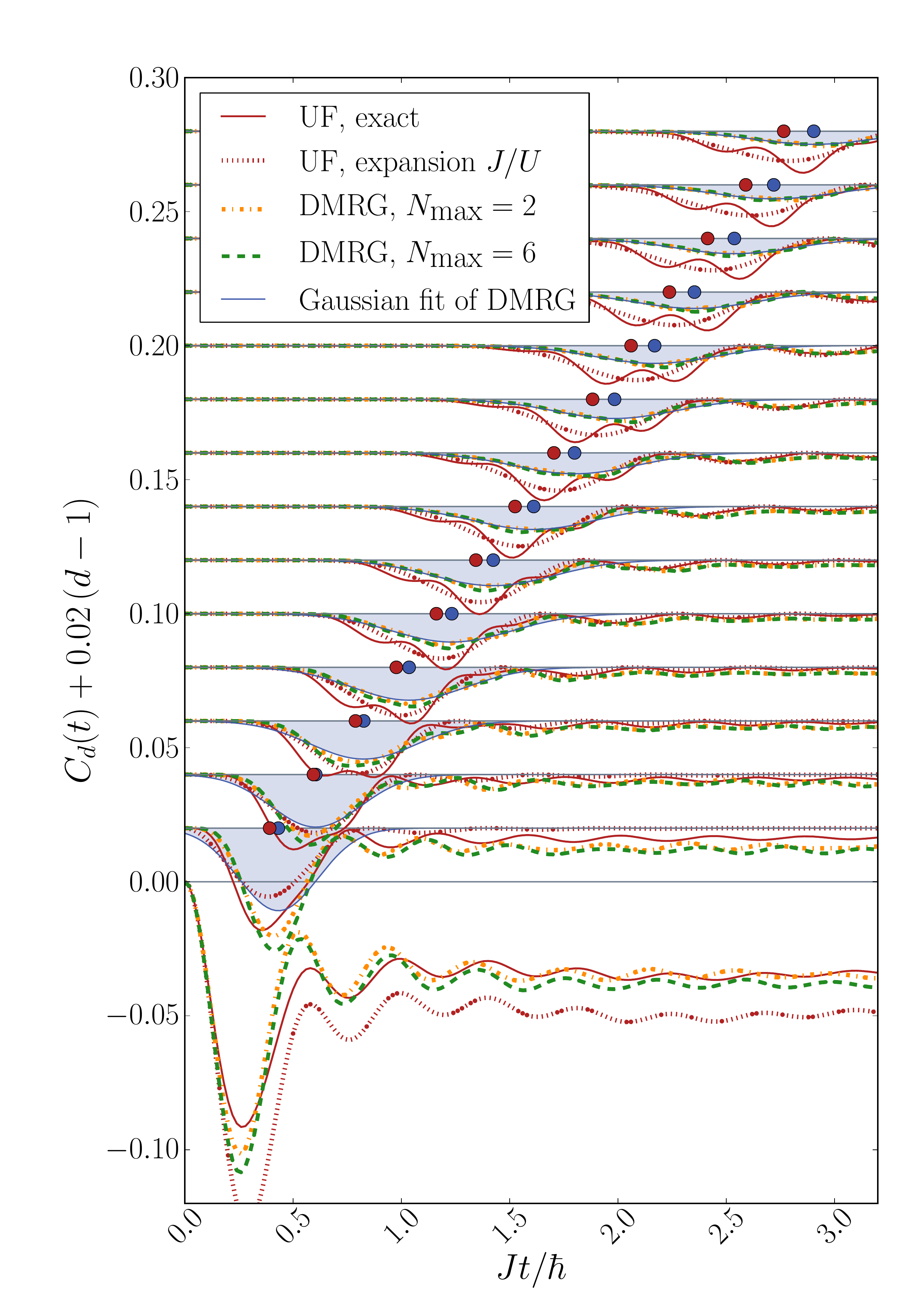}
    \end{center}
    \caption{ \label{fig:quenchcorr2}\figpreamble Dynamics of density correlations after a quench from the Fock
      state $\ket{\nnot}$ at infinite interactions to a final interaction $U/J=9$. See
      Fig.~\ref{fig:quenchcorr1} for more information.}
  \end{figure}

  For quenches to intermediate values of the interaction strength, the dynamics of density
  correlations exhibits essentially the same features as described above. This can be seen in
  Fig.~\ref{fig:quenchcorr2}, where the dynamics following a quench to $U/J=9$ is depicted.  In
  particular, the characteristic dip corresponding to the propagating quasiparticles is still
  present. We note that the propagation of this correlation signal is still remarkably well
  described by the UF approximation, even though this model is close to breaking down at this
  interaction strength. However, one sees that the strong coupling expansion significantly
  overestimates the amplitude of the correlations at $d=1$ and that the amplitude of the unphysical
  oscillations in the numerical evaluation of the UF approximation increases.

\begin{figure}[ht]
    \begin{center}
      \includegraphics[width=0.49\textwidth]{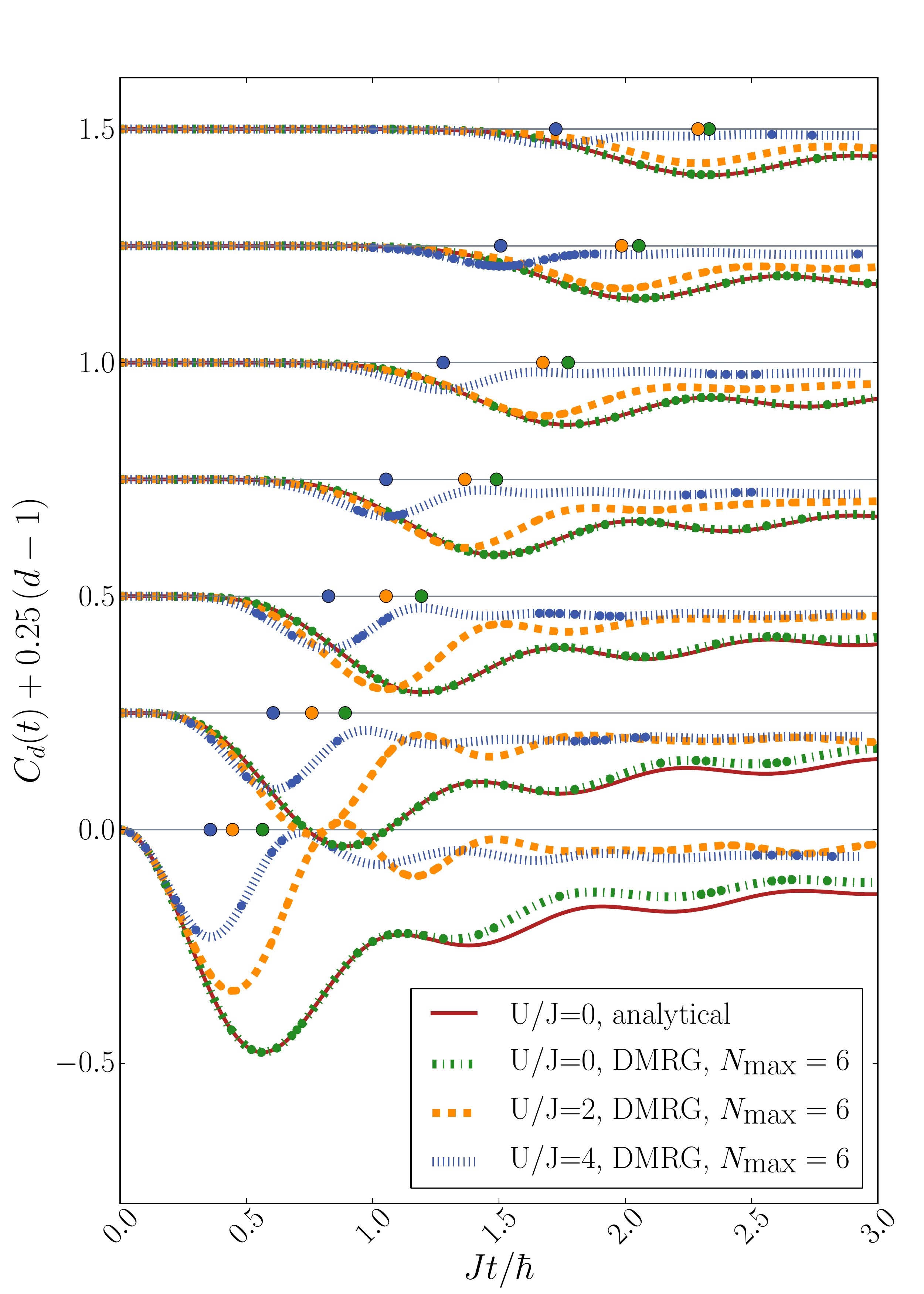}
    \end{center}
    \caption{\label{fig:quenchcorr3}\figpreamble Dynamics of density correlations after a quench
      from a Fock state $\ket{\nnot}$ at infinite interactions to weak final interactions. The
      results for different distances $d$ are shifted for clarity by $0.25(d-1)$. Unlike in
      Figs.~\ref{fig:quenchcorr1} and \ref{fig:quenchcorr2}, the position of the correlation signal of the DMRG results is
      identified with the position of the absolute minimum and it is denoted by the circles in the corresponding colors.}
  \end{figure}

  In the weakly interacting regime, one could expect a different behavior since the relevant
  quasiparticles are of different nature. However, the main features that characterize the dynamics
  of density correlations at strong and intermediate interactions are remarkably preserved,
  as can be seen in Fig.~\ref{fig:quenchcorr3}. In particular, a propagating dip can be identified in
  all cases. The main difference between the non-interacting \eqref{eq:corrnonint} and the strongly
  interacting \eqref{eq:corrapprox} cases is the lower velocity and very slow decay of correlations
  at long times when $U=0$. At $U/J=2$, this long tail is already strongly suppressed at short
  distances, but is still visible at longer distances. For $U/J=4$ (Fig.~\ref{fig:quenchcorr3}), one
  sees that the overall profile of the propagating correlation signal is already very similar to that in the
  more strongly interacting case (Figs. \ref{fig:quenchcorr1} and \ref{fig:quenchcorr2}).

  At low interaction strengths, one has to be careful when using the DMRG simulations since the
  truncation of the local Hilbert space to a finite number of states can introduce significant
  errors. By comparing DMRG simulations to the exact formula \eqref{eq:corrnonint} obtained in the
  ``worst'' case $U/J=0$, we found that a maximal site occupancy $N_{\text{max}}=6$ represents a
  fairly safe approximation, whose accuracy improves with the distance for the times considered (see
  Fig.~\ref{fig:quenchcorr3}).

  To summarize the analysis conducted in this section, we observe that the dynamics of the density
  correlations is dominated by the propagation of a negative signal (dip). For strong interactions,
  this dip results from the propagation of the quasiparticle pairs described in Sec.
  \ref{sec:prop-quas}. In the next section we will investigate in more detail the shape of that
  signal and characterize its propagation velocity quantitatively.


  \begin{figure*}[ht]
    \begin{center}
      \includegraphics[width=\textwidth]{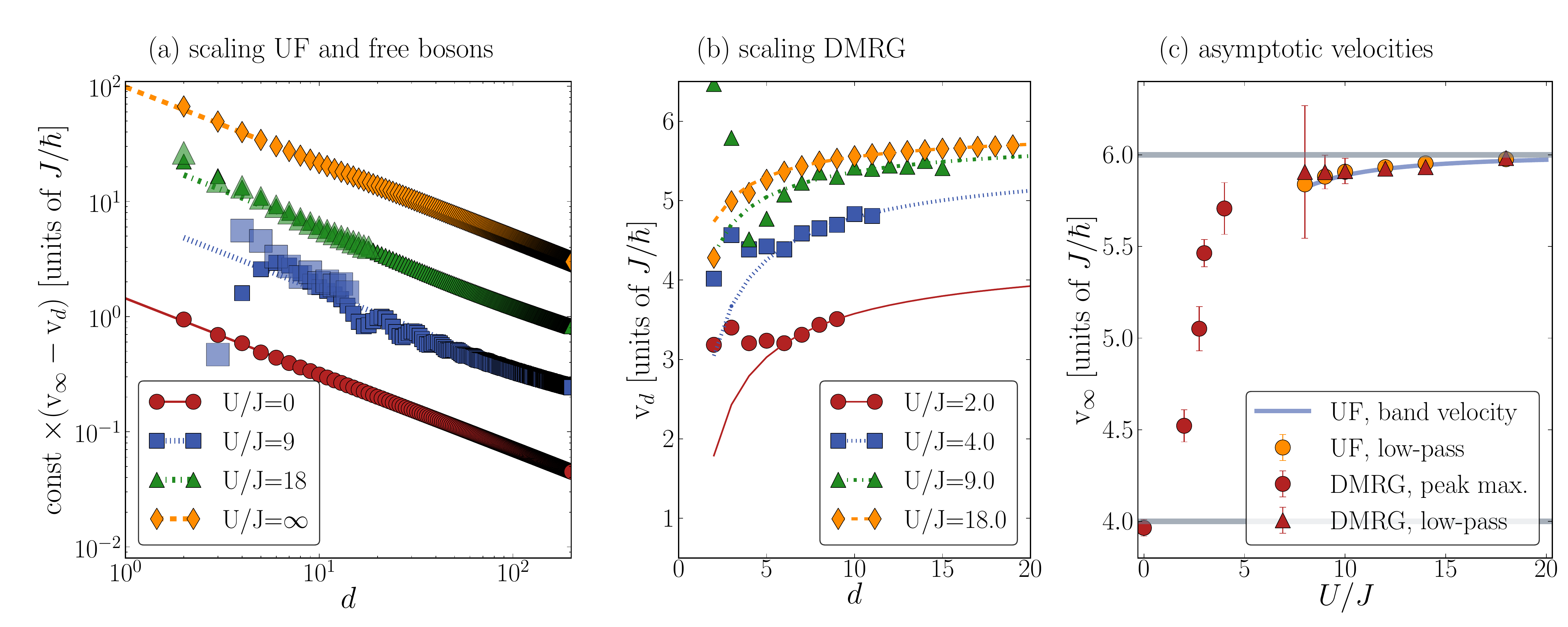}
    \end{center}
    \caption{\label{fig:v}\figpreamble (a) Instantaneous propagation velocity in the UF approximation or in the
      non-interacting case as a function of the distance $d$ (filled symbols). Light symbols
      represent DMRG data. Lines show the fits $|\text{v}_d-\text{v}_\infty|\propto
      d^{-\alpha}$. The data have been shifted vertically for a better visibility. (b) Instantaneous
      propagation velocity obtained by DMRG simulation as a function of the distance $d$ (filled
      symbols). Lines show the fits $|\text{v}_d-\text{v}_\infty|\propto d^{-\alpha}$ with fixed
      exponent $\alpha=0.65$. (c) Asymptotic velocities $\text{v}_\infty$ extracted from the finite
      distance data versus interaction strength using $|\text{v}_d-\text{v}_\infty|\propto
      d^{-0.65}$. Error bars denote the 2-sigma uncertainty of the fit that yields the asymptotic
      velocity.  }
  \end{figure*}

  \subsection{Analysis of the signal propagation}
  \label{sec:corrpropagation}
  We can get a lot of insight into the propagation of the signal from the following approximation of
  the density correlations at large distances (we recall that the lattice constant $a_{\text{lat}}$
  is set to one):
  \begin{equation}
    C_{d} \stackrel{d \gg 1}{\approx} -\left(\frac{2d^{2/3}2^{1/3}\hbar}{3Ut}\right)^2
    \text{Ai}^{\,2}\left(-(2/d)^{1/3}(6Jt/\hbar-d)\right)\,.\label{eq:corrairy}
  \end{equation}
  In the above expression, derived from (\ref{eq:corrapprox2}), we made use of the relation existing
  between the Airy function $\text{Ai}(-z)$ and the high-order Bessel functions
  \cite{AbramovizBook}:
  \begin{equation}
    \Jb_d(d+z d^{1/3}) =2^{1/3} d^{-1/3} {\rm Ai}\big( -2^{1/3} z\big) + \mathcal{O}\big(d^{-1}\big)\,.
  \end{equation}
  The Airy function is plotted in the inset of Fig.~\ref{fig:quenchcorr1}. It
  exhibits a peak located at $z_0\approx 1.02$ and surrounded by an exponential tail on the side
  $z<z_{0}$ and by an algebraically-decaying oscillating tail on the side $z> z_0$.

  Disregarding the monotonically and slowly varying prefactor in Eq.~\eqref{eq:corrairy}, the profile
  of the Airy function alone allows us to understand several features of the propagation of the
  correlation signal. For example, it reveals the existence of a well defined propagation front,
  since the correlations are exponentially suppressed for times $t<t_{\text{peak}}$, with
  \begin{equation}
    \frac{Jt_{\text{peak}}}{\hbar} \approx \frac{1}{6} \left[ d + z_0 \left( \frac{d}{2} \right)^{1/3} \right] \,.
  \end{equation}
  The signal in the density correlations corresponds to the peak of the Airy function. Once this
  peak has passed, that is for $t>t_{\text{peak}}$, the correlations show an algebraic decay with
  oscillations. From the definition of $t_{\text{peak}}$, one sees that two terms contribute to the
  propagation of the correlation signal: the first is simply proportional to the distance,
  corresponding to a well defined velocity, whereas the second is proportional to $d^{1/3}$.
  The linear contribution dominates at large distances, leading to a light-cone-like
  spreading of the correlations. At small distances, however, the dynamics deviates significantly
  from the asymptotic light cone. This behavior can be accounted for by defining an 'instantaneous'
  propagation velocity:
  \begin{align}
    \text{v}_d & =  \left[ t_{\text{peak}}(d+1)-t_{\text{peak}}(d) \right]^{-1} \notag\\
    & = \text{v}_\infty\left( 1-\frac{z_0}{2^{1/3} 3} d^{-2/3} \right) + \mathcal{O}\big(d^{-5/3}
    \big)\,.\label{eq:vscaling}
  \end{align}
  One sees immediately in the above equation that the asymptotic light cone is characterized by the
  velocity $\text{v}_\infty=6J/\hbar$ and is reached algebraically at large distances, whereas the
  propagation velocity can go down to approximately $4J/\hbar$ at short distances.

  A similar analysis can be carried out for the case $U/J=0$. It turns out that the correlation dip
  is almost completely described by a single term in the infinite sum (\ref{eq:corrnonint}). For
  even $d$, for example, we obtain:
  \begin{align}
    C_{d}(t) & \approx-2\Jb^4_{d/2}(2Jt/\hbar)\notag\\
    & = -2d^{-4/3}\text{Ai}^4\left(-d^{-1/3}\left(4Jt/\hbar-d\right)/2\right)
    \,.\label{eq:corrairyni}
  \end{align}
  The same expression for the instantaneous velocity (\ref{eq:vscaling}) therefore holds in the
  non-interacting case as well, but with $\text{v}_\infty=4J/\hbar$, which is the velocity of freely
  propagating bosons. The behavior at $U/J=0$ mostly differs from the strongly interacting case
  once the correlation dip has passed ($t>t_{\text{peak}}$): further terms (\ref{eq:corrnonint})
  beyond Eq.~\eqref{eq:corrairyni} then become important, which causes correlations to decay very
  slowly (see Fig.~\ref{fig:quenchcorr3}).

  The width and the height of the correlation dip can also be derived from the expressions
  (\ref{eq:corrairy},\ref{eq:corrairyni}). For both the interacting and the non-interacting case,
  the width increases proportionally to $d^{1/3}$ while the height decreases with $d^{-2/3}J^2/U^2$
  in the strongly interacting case and with $d^{-4/3}$ for $U/J=0$. We note that similar power laws
  have been found for the quantum Ising model \cite{Igloi2000}.

  In the following, we show that the approximate scaling of the velocity found in the strongly and
  non-interacting limits holds for any interaction strength. We first concentrate on large
  interaction strengths.  Within the UF approximation, we can evaluate the correlations up to
  arbitrarily long times and make a rigorous scaling analysis of the instantaneous propagation
  velocity. We determine the position of the dip by means of a Gaussian fit after having filtered
  out oscillations with a period shorter than $h/U$ using a low-pass filter. Fig.~\ref{fig:v}(a)
  illustrates for a few interaction strengths that the analytical scaling behavior
  $|\text{v}_d-\text{v}_\infty|\propto d^{-\alpha}$ is accurately reproduced at sufficiently large
  distances $d>5$. Extracting the asymptotic velocities $\text{v}_\infty$ and the exponents $\alpha$
  with a fit over distances $6\leq d\leq 400$, we obtain values in very good agreement with the
  approximated analytical predictions. For example, the exponent is found to be the same for all
  interactions: $\alpha=0.650\pm 0.002$. The small difference from the value $\alpha=2/3$ expected
  from the Airy functions \eqref{eq:vscaling} is most probably due to the prefactor in
  (\ref{eq:corrairy}), which we neglected when deriving \eqref{eq:vscaling}. The asymptotic
  velocities match the ones that we expect from the quasiparticle dispersion relation
  Eq.~\eqref{eq:vmax}, as shown in Fig.~\ref{fig:v}(c). Close to the breakdown of the UF
  approximation, the oscillation frequencies due to the interaction and the finite bandwidth become
  similar and one cannot easily filter out the first one anymore. The instantaneous velocity
  $\text{v}_d$ therefore shows an oscillatory behavior even at very large distances $d\lesssim
  50$. Nevertheless, the scaling behavior remains perfectly obeyed on average and in the
  long-distance limit. In the non-interacting case, shown in Fig.~\ref{fig:v}(a), we extract
  accurately the position of the correlation signal by simply locating the first minimum. We again
  find the scaling exponent $\alpha=0.650\pm 0.002$ and the extracted asymptotic velocity is close
  to the expected value $\text{v}_\infty=4J/\hbar$ [cf.~Fig.~\ref{fig:v}(c)].

  Using DMRG simulations, we can calculate the dynamics exactly for all interaction strengths, but we
  are restricted to short time and length scales. We therefore fix the scaling exponent to
  $\alpha=0.650$ in order to extract the asymptotic velocities.  In Fig.~\ref{fig:v}(b) we show that
  the scaling $|\text{v}_d-\text{v}_\infty|\propto d^{-0.65}$ becomes accurate as the distance
  increases for both strong ($U/J\geq8$, extracted with low-pass filter and Gaussian fit) and weak
  interactions ($U/J\leq4$ extracted directly from the peak without low-pass filter).  Despite the
  limited number of data points available in the scaling region, we can determine the asymptotic
  velocities with a reasonably small uncertainty. The values that we obtain, gathered in
  Fig.~\ref{fig:v}(c), are in good agreement with those predicted by the UF approximation. The lack
  of data in the range $4<U/J<8$ results from the mixing of the time scales related to kinetic and
  interaction processes and which prevents us from locating accurately
  the position of the correlation signal. The asymptotic velocities in Fig.~\ref{fig:v}(c) can be
  seen as a characterization of a crossover between a regime of quasi-free bosons ($U/J\lesssim4$),
  with a renormalized velocity, and the strongly interacting regime described by two flavors of fermions. This crossover is not
  directly related to the ground-state phase diagram of the Bose-Hubbard model since the propagation
  velocity reflects the dispersion in the center of the Brillouin zone (at wave vectors
  $k\approx\pm\frac{\pi}{2}$), rather than low-wavelength modes. As a consequence, $\text{v}_\infty$
  is considerable higher than the sound velocity in the superfluid regime \cite{Lieb1963,Kollath2005} and a
  linear propagation with $\text{v}_\infty\lesssim 6J/\hbar$ is found at strong interactions, even
  though at equilibrium the system would be in the Mott-insulating phase.
  
  As a final remark, we note that the dependency of the propagation velocity on $U/J$ in that system
  has been studied before by L\"auchli and Kollath \cite{Lauchli2008}, who considered the case of a
  quench from a small interaction strength to a larger one. Surprisingly, the instantaneous
  spreading velocity has been found to exhibit a maximum at intermediate interaction strength. A
  possible explanation for this effect could be that bosonic atom number fluctuations present in the
  initial superfluid state may lead to enhanced velocites as compared to the quench from the Fock
  state. A quantitative comparison between our predictions and this previous work would require an
  extrapolation of the velocity to large distances which is difficult in the absence of an
  analytical model.

  \section{Conclusions}
  \label{sec:conclusions}
  In order to describe accurately the quench dynamics of the one-dimensional Bose--Hubbard model in
  the Mott-insulating regime, we have developed a new analytical approach relying on the
  fermionization of auxiliary bosons. Its predictions regarding both the ground state and the
  dynamical properties are found in quantitative agreement with exact numerical simulations for
  large and intermediate interaction strengths $U/J > 8$. This constitutes a great improvement with
  respect to the analytical models introduced previously.

  Using this model, we are able to investigate the time evolution of density correlations in the
  quenched system over exceedingly long times. We observe a characteristic light c/one dynamics,
  meaning that there exists a distance, linearly growing in time, beyond which correlations 
  between distant sites are exponentially suppressed.
  More precisely, correlations spread as a wave packet along this light cone, forming a
  propagation front whose position can be unambiguously identified. A careful analysis of the
  velocity with which this front propagates reveals a generic scaling behavior characterized by a
  universal exponent and an asymptotic velocity dependent on the interaction strength. The same
  behavior is found in the non-interacting limit of freely propagating bosons, where an explicit
  solution is available, as well as in the intermediate regime $0 < U/J \leq 8$, where we
  rely on exact numerical simulations. The asymptotic velocity, which varies significantly
  between the weakly and the strongly interacting regime, is a useful quantity to characterize a broad 
  spectral range of the Hamiltonian as it does not depend only on its low-lying modes.

  Building upon this first success, we envisage that the representation of the Bose--Hubbard model
  in terms of fermionic quasiparticles could shed new light on the mechanism for thermalization
  or serve as a tool to interpret the outcome of spectroscopic measurement on laboratory
  systems, such as modulation spectroscopy or Bragg spectroscopy for ultracold gases in optical
  lattices.

\begin{acknowledgements}
  We thank D. Baeriswyl, T. Giamarchi, V.  Gritsev, S. Huber and A. Tokuno for
  discussions. Financial support by ANR (FAMOUS), SNSF under Division II and MaNEP, and EU (Marie
  Curie Fellowship to M.C.) is acknowledged. P.B. thanks the University of Fribourg for hospitality.
\end{acknowledgements}

\appendix
\section{Perturbation theory}
\label{sec:perturbation}
In this appendix we develop a complementary perturbative approach to recover the behavior in the
strong coupling limit to first non-vanishing order in $J/U$. The situation we consider is the quench
from the Fock state $|\psi(0)\rangle=\fockket$ at filling $\nnot$ to a large final
interaction strength $U/J$.

In all generality, the wave function after a sudden change of parameters can be written in the
eigenbasis $\ket{\phi_n}$ (with corresponding eigenenergies $E_n$) of the final Hamiltonian
\begin{equation}
  | \psi(t) \rangle = \sum_n e^{-it \frac{E_n}{\hbar}} \langle \phi_n | \psi(0) \rangle\, |\phi_n\rangle\,.
\end{equation} 
 Usually the difficulty lies in determining the
eigenstates $\ket{\phi_n}$ and their corresponding energies $E_n$ in a many-body problem. In this
appendix we determine $\ket{\phi_n}$ and $E_n$ by perturbation theory in $J/U$ in a system of length
$L$ with periodic boundary conditions. Note, that this is not a full perturbative expansion, since we
will not expand the exponential in the corresponding power.

We consider the interaction term of the Bose--Hubbard Hamiltonian as the unperturbed Hamiltonian and
the kinetic part as the perturbation. The eigenenergies of the unperturbed Hamiltonian are multiples
of the interaction strength $U$ and the corresponding states are Fock states. More precisely, the
ground state is the Fock state $\fockket$ with vanishing energy. The lowest excited
states are the states with a single particle-hole excitation with energy $U$. These we denote by
$|\phi(m,d)\rangle$ with an occupation $\nnot$ for all the sites except for site $m$ with $\nnot+1$
atoms and the site $m+d$ with $\nnot-1$ atoms, i.e.~$|\phi(m,d)\rangle=
\frac{1}{\sqrt{\nnot(\nnot+1)}} b_{m+d}b_m^\dagger \fockket$.  Using degenerate perturbation theory
(restricted to the same symmetry sector as the initial state) at first order in $J/U$, the ground
state energy remains zero and the ground state of the final Hamiltonian is given by
\begin{align}
  |\phi_0\rangle \approx \fockket + \frac{\sqrt{\nnot(\nnot+1)}J}{ U} \sum_{m=0}^{L-1}\left(
    |\phi(m,1)\rangle + |\phi(m,-1)\rangle \right).
\end{align}
The lowest band of excited states resulting from the single particle-hole excitations is formed by
\begin{align}
  |\phi_p\rangle \approx |\phi_p^0\rangle - \frac{\sqrt{2\nnot(\nnot+1)}J}{ U} \eta_p \sin(\pi p
  /L)\fockket +J/U \sum_\alpha\ket{\tilde{\phi_\alpha}} \label{eq:perturbedstates}
\end{align}
with corresponding energies
\begin{align}
  E_p \approx U-2(2\nnot+1)J\cos(\pi p /L)\,.
\end{align}
Here $ |\phi_p^0\rangle$ ($p=0,\dots,L-1$) are the symmetric states that diagonalize the kinetic
part of the Hamiltonian given by $$ |\phi_p^0\rangle = \frac{\sqrt{2}}{L} \sum_{d=1}^{L-1}
\sum_{m=0}^{L-1} \sin(\pi p d / L) |\phi(m,d)\rangle\,.$$ Note that the index $d$ only starts at $1$
to avoid the double counting of the Fock state. We employed the notation
$\eta_p=(1-(-1)^p)$.  As we are interested in the time-evolution of the initial Fock
  state, we abbreviated unimportant terms as $\ket{\tilde{\phi_\alpha}}$ which are the states
beside the Fock state that are directly coupled via the kinetic term to the states
$|\phi_p^0\rangle$.

Using these eigenenergies and states to first order, we can now write the time evolving state $|
\psi(t) \rangle$ as
\begin{multline}
  |\psi (t)\rangle = \fockket + \frac{J\sqrt{\nnot(\nnot +1)}}{U } \sum_m\left( |\phi(m,1)\rangle +
    |\phi(m,-1)\rangle \right) \\
  - \frac{\sqrt{2\nnot(\nnot+1)}J}{U}  \sum_p \eta_p \sin(\pi p /L) e^{-i \frac{ E_p}{\hbar} t }
  |\phi_p^0\rangle \,.
\end{multline}
This formula corresponds to the expression (\ref{eq:psipert}) which one obtains in the unconstrained
fermionic approach up to first order. As discussed in the unconstrained fermionic approach this
expression gives a lot of insight into the formation and propagation of singly and doubly occupied
sites and can be used to compute all the observables that we are interested in.

%

\end{document}